\documentclass[sort&compress,AMA,STIX1COL]{WileyNJD-v2}

\usepackage{hyperref}
\usepackage{mathtools}
\usepackage{amssymb}
\usepackage{amsmath}
\usepackage{graphicx}
\usepackage[caption=false]{subfig}

\usepackage{xcolor}

\articletype{Research Article}%

\received{xxx}
\revised{xxx}
\accepted{xxx}

\begin{document}
	
	\title{Exploiting non-orthogonal multiple access in downlink coordinated multipoint transmission with the presence of imperfect channel state information}


\author[1,3]{Fahri Wisnu Murti}

\author[1,3]{Rahmat Faddli Siregar}
\author[2,3]{Muhammad Royyan}

\author[3]{Soo Young Shin*}


\address[1]{\orgdiv{Faculty of Information Technology and Electrical Engineering (ITEE)}, \orgname{University of Oulu}, \orgaddress{\country{Finland}}}

\address[2]{\orgdiv{Department of Signal Processing and Acoustics}, \orgname{Aalto University}, \orgaddress{\country{Finland}}}

\address[3]{\orgdiv{Dept. of IT Convergence Engineering}, \orgname{Kumoh National Institute of Technology}, \orgaddress{\country{South Korea}}}


\corres{*Soo Young Shin. \email{wdragon@kumoh.ac.kr}}

\presentaddress{Dept. of IT Convergence Engineering, Kumoh National Institute of Technology (KIT), 39177, Gumi, South Korea.}

\abstract[Summary]{
	\textcolor{blue}{In this paper, the impact of imperfect channel state information (CSI) on a downlink coordinated multipoint (CoMP) transmission system with non-orthogonal multiple access (NOMA) is investigated since perfect knowledge of a channel can not be guaranteed in practice. Furthermore, the channel estimation error is applied to estimate the channel information wherein its priori of variance is assumed to be known. The impact of the number of coordinated base stations (BSs) on downlink CoMP NOMA is investigated. Users are classified into one of two groups according to their position within the cell, namely cell-center user (CCU) and cell-edge user (CEU). In this paper, ergodic capacity and sum capacity for both CCU and CEU are derived as closed form. In addition, various experiments are conducted with different parameters such as SNR, error variance, and power allocation to show their impact on the CoMP method. The results show that CoMP NOMA outperforms the CoMP orthogonal multiple access (OMA) wherein the condition of the channel impacts the performance of CoMP NOMA less. It is worth noting that a higher number of coordinated BSs enhances the total capacity of CoMP NOMA. Finally, the performance analysis is validated due to the close accordance between the analytical and simulation results.} }




\maketitle

\section{Introduction}\label{sec1}
The upturn trend in multimedia applications and massive connections such as Internet of Things (IoT), require enormous data traffic, which motivates further research for the next generation wireless technology, i.e., 5G to fulfill the capacity demand. Compared to 4G, 10 to 100 data rate enhancements are demanded for 5G \cite{3gpp}. To fulfill this requirement, researchers come together to design the framework and standardization of 5G \cite{5g_wu,5g_ntt,5g_nakamura,5g_shafi}. Among all of the 5G candidates, non-orthogonal multiple access (NOMA) has become a potential candidate due to its high spectral efficiency \cite{zding_impact,basit1,murti_cqi,murti_mimo}. In addition to the benefit of high spectral efficiency, NOMA has been considered as essential scheme to support massive connectivity such as IoT, that can not be effectively supported by orthogonal multiple access (OMA) scheme  \cite{noma_iot,zding_survey}. 

In NOMA, two or more users are paired/grouped together and allocated different power within the same resource, which can be of the frequency, time, or spreading code. Further, the signals are superimposed and multiplexed into the power domain. In downlink NOMA, the interference cancellation techniques such as successive interference cancellation (SIC) is performed at the high channel gain user to remove the signals from the lower channel gain users. At lowest channel gain user, the user does not perform SIC and decodes its own signal by considering the higher channel gain users as noise \cite{zding_impact,basit1}. By sharing the same resource, either each user or/and the overall system can obtain significant capacity. In NOMA, power allocation and user grouping scheme play an important role to significant capacity improvement, otherwise, an insufficient scheme can also lead to performance degradation \cite{zding_impact,basit1,murti_cqi,murti_mimo}. 

Most of research on NOMA was focused on the single-cell NOMA, while multi-cell scenario did not get enough attention \cite{shin_comp}. That is due to the fact that multi-cell generate interference from each adjacent cell that could lead to performance degradation \cite{ali_large}. Hence, interference management of multi-cell scenario is required to be investigated. The conventional scheme, i.e., coordinated multi-point (CoMP) transmission, has been known as the critical enhancement for LTE Advance \cite{sawashi,irmer}. In CoMP, users that are covered by multiple base stations (BSs), are served by multiple coordinated BSs simultaneously to mitigate the inter-cell interference which lead to improving cell-edge user (CEU) performance. However, in OMA, all the involved BSs need to allocate the same resource to the CoMP users exclusively, which cannot be utilized by the other users. Consequently, when the number of CoMP users increases, the spectral efficiency becomes significantly worse \cite{sun_3}.

To overcome the CoMP OMA problem, NOMA scheme is utilized in the CoMP scenario \cite{sun_3,choi_compnoma, ali_compnoma,tian_oppor}. Several studies have been conducted to combine CoMP with NOMA. In \cite{choi_compnoma}, the Alamouti code is applied to the two coordinated BSs for the joint transmission scenario. NOMA is also utilized by grouping two cell-center users (CCUs) and a single CEU. In this scenario, all the CCUs act as a non-CoMP user who receives the interference from the coordinated neighboring BS, while CEU act as a CoMP user. Additionally, multiple scenarios involving the coordinated two-points system are also studied in \cite{ali_compnoma} by employing the coordinated beamforming and joint transmission scheme. The authors also compared the proposed CoMP NOMA scenario with the traditional CoMP OMA.  In \cite{sun_3},  CoMP NOMA with the three-points system is investigated with a randomly deployed user. Distributed analog beamforming is also applied at the transmitter to meet the CEU quality of service requirements. However, the authors in \cite{sun_3,choi_compnoma,ali_compnoma} only only focus on a specific number of BSs such as two-points or three-points system, instead of with $B$ number of total BSs.

Recently, the a generalized opportunistic NOMA (ONOMA) in the CoMP system has been proposed in \cite{tian_oppor}. ONOMA was proposed to improve the capacity and reduce the complexity of CoMP NOMA. The mathematical analysis was also provided for ONOMA. User association and power allocation schemes are also proposed in \cite{compnoma_pa,ali_compnoma2}. However, these studies \cite{sun_3,choi_compnoma, ali_compnoma,tian_oppor, compnoma_pa,ali_compnoma2}, only include an analysis on perfect channel state information (CSI). In practical studies, the assumption of perfect CSI at the transmitter is impractical since obtaining perfect CSI requires enormous consumption of the system overhead \cite{yang_par}. Moreover, in CoMP NOMA, many of users and multiple number of coordinated BSs are involved which cause a higher system overhead consumption than the general NOMA scenario; and serving high-mobility users, such as high speed trains and cars is required in the future wireless network \cite{3gpp,yang_par}. Despite some studies have considered imperfect CSI in NOMA \cite{yang_par,fangfang,ygao}; however, these works did not include the coordination between multi-cell networks in the imperfect CSI analysis. Therefore, the study on CoMP NOMA, considering that CSI cannot be obtained correctly, is also compulsory to design 5G frameworks.  

 Our work contributes to CoMP NOMA design from a new angle, offers a rigorous analysis of ergodic capacity, highlights the importance study of imperfect CSI.  The overall contributions of this paper are listed as follows:
\begin{itemize}
	\item[$\bullet$] Due to the utilization of the NOMA scheme, two types of pairing scenarios are considered. Firstly, CEU is paired among CCU with affected interference from the other coordinated BSs. In this case, either CCU or CEU will receive the signal from the other BSs. Otherwise, the second type of scenario, the CCU is only influenced from the main serving BS. For both scenarios, each CoMP NOMA pair/group consist of $B+1$ users that share the same resource simultaneously.
	\item[$\bullet$] The CoMP NOMA system is investigated if CSI cannot be obtained perfectly. Imperfect CSI is modeled with the presence of channel estimation error where a priori of the variance of the error estimation is known. Moreover, instead of assuming perfect SIC as in \cite{yang_par}, the imperfectness of SIC process is also considered, owing to sequel effect of estimation error. 
	\item[$\bullet$] The closed-form solution for the exact ergodic capacity of CoMP NOMA in the presence of channel estimation error is derived over independent Rayleigh fading channel. The closed-form solution can also be used to calculate the analytical result in case of perfect CSI can be obtained.
	\item[$\bullet$] Finally, the simulation and analytical results are presented to provide the detailed performance of CoMP NOMA, with and without the presence of imperfect CSI. The analytical and simulation results are shown simultaneously to validate the correctness of the performance analysis.
\end{itemize} 

The rest of this paper is organized as follows: Section \ref{sec_sys} presents the system and channel model of CoMP NOMA including perfect CSI and channel estimation error. The closed-form solutions for the exact ergodic capacity of CCUs, CEUs, and sum capacity are presented in Section \ref{sec_performance}. Then, Section \ref{sec_result} provides the detailed performance evaluation of CoMP NOMA in the presence of channel estimation error. The closed-form solutions are also validated through the analytical and simulation results in this section \ref{sec_result}. Finally, Section \ref{sec_conclusion} conclude of overall this paper.
\section{System and channel model} \label{sec_sys}
In this section, the system model of $B$ coordinated BSs of CoMP NOMA with perfect and imperfect CSI is presented. For imperfect CSI, channel estimation error is modeled.
\subsection{CoMP NOMA with perfect CSI}
\begin{figure}[!t]
	\centering
	\subfloat[The system model of CoMP NOMA with $B$ coordinated BSs within a single group $G_k$ with same frequency resource] {\label{subfig:system2}\includegraphics [width=0.45\textwidth]{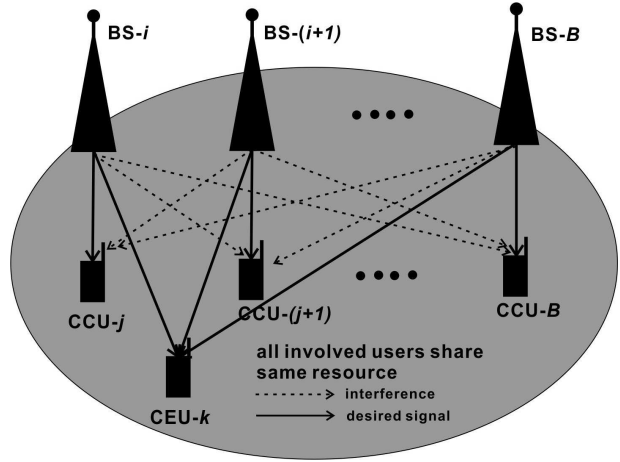}}   \\ 
	\subfloat[The observed CoMP NOMA group $G_1$ with $B=3$ ] {\label{subfig:system1}\includegraphics [width=0.45\textwidth]{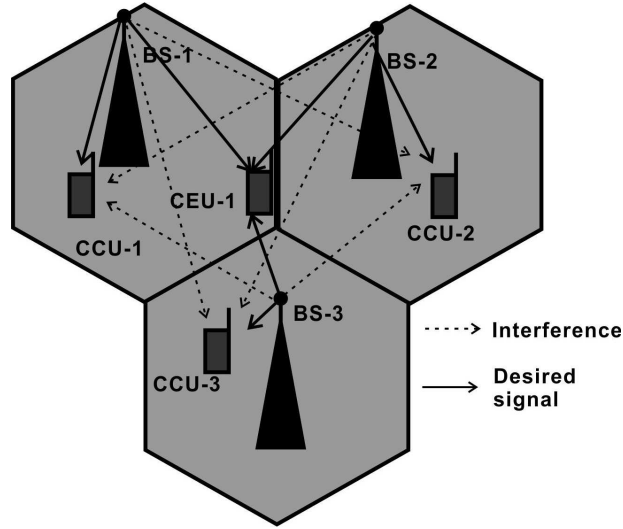}}    
	\caption {System model of CoMP NOMA within a single group}
	\label{fig.system}
\end{figure}

In this system, the power-based NOMA is applied to CoMP scheme. In NOMA, the users share same resource, which can be a time slot, a frequency resource, or a spreading code \cite{basit1,murti_cqi}. Consider the downlink coordinated multiple cells scenario with total $B$ multiple BSs communicating with CCUs and CEUs simultaneously through the CoMP NOMA scenario. Let us suppose $G_k$, $1 \leq k \leq K$, represents the $k$-th CoMP NOMA user grouping, which consists of the selected multiple CCUs and the single CEU. The selected single CEU of $G_k$ is denoted with CEU-$k$, $1 \leq k \leq K$. For simplicity, we assume that a single BS only serves a single CCU. Therefore,  the $i$-th user of CCUs which near the coordinated BS-$i$ is denoted with CCU-$i$, $1 \leq i \leq B $. Hence, a single CoMP NOMA group, consists of total $B$ multiple CCUs and a single CEU as shown in Figure \ref{fig.system}. This user grouping scheme is applied to maintain the channel gain difference of the NOMA scheme. For instance, the observed CoMP user CEU-$1$ of $G_1$ performs the CoMP NOMA scheme with coordinated BS-1, BS-2 and BS-3. The CEU-1 will share same resource with CCU-1, CCU-2, and CCU-3 within the CoMP NOMA group $G_1$ as shown in Figure \ref{subfig:system1} . Consequently, the channel gain difference between each user can be maintained. Note that maintaining the channel gain difference between paired/grouped users is important factor to maximize the capacity in the NOMA scheme \cite{zding_impact,basit1, murti_cqi}.  

For each BS-$i$, the superposition code is applied by the following NOMA scheme. Therefore, BS-$i$ transmits the following signal information
\begin{equation} \label{eq.tx}
x_i = \sqrt{\alpha_{i}} s_i +  \sqrt{\alpha_{k}} s_k,
\end{equation}
where $s_i$ represents the desired signal for CCU-$i$ from BS-$i$, $s_k$ represents desired signal for CEU-$k$, and $\alpha_i,\alpha_k$ are the normalized power allocation to $i$-th and $k$-th user with conditions $\alpha_i+\alpha_k=1$ and $\alpha_i \leq \alpha_k$ , respectively. For simplicity, we assume that $\alpha_i$ and $\alpha_k$ are determined to be constant to all BS as $\alpha$ and $\beta$ \cite{sun_3}. 
In this system, the analysis is divided into two parts: CCU as the non-CoMP users and CEU as the CoMP users.
\subsubsection{Cell-center user}
In this scenario, the CCU-$i$ is only served by BS-$i$, which means CCU-$i$  does not perform the CoMP scheme with the other BSs. Further, the received signal of the observed CCU-$j$, $1 \leq j \leq B$ can be expressed as follows
\begin{equation} \label{eq.rxj}
\begin{aligned}
y_j &= \sum_{i=1}^{B}  h_{ij}\left( \sqrt{\alpha} s_i + \sqrt{\beta} s_k \right) + n_j \\
&= h_{jj}\left( \sqrt{\alpha} s_j + \sqrt{\beta} s_k \right)
 + \sum\limits_{\substack{i=1 \\ i \neq j}}^{B} \underbrace{h_{ij}\left( \sqrt{\alpha} s_i + \sqrt{\beta} s_k \right)}_\textrm{interference from the other BSs} + n_j,
\end{aligned}
\end{equation}
where $h_{ij} \sim CN\left(0,\sigma_{ij}^2 = d_{ij}^{-v}  \right)$ is denoted as channel coefficient from BS-$i$ to observed CCU-$j$ which follows
independent and identically distributed (i.i.d) Rayleigh fading, $d_{ij}$ is distance normalization between transmit antenna of BS-$i$ to observed CCU-$j$, $v$ is path loss exponent parameter, and $n_j \sim CN(0,1)$ represents noise at CCU-$j$. Further, SIC is performed at all the selected CCUs to remove the CEU signal before decoding its own signal. In case the CEU signal is perfectly removed, the signal-to-interference noise ratio (SINR) of CCU-$j$ can be calculated as follows
\begin{equation} \label{eq.sinrj}
\gamma_j = \frac{\alpha |h_{jj}|^2}{\sum\limits_{\substack{i=1 \\ i \neq j}}^{B} \alpha |h_{ij}|^2 + \frac{1}{\rho}},
\end{equation}
where $\rho$ represents the transmit signal-to-noise ratio (SNR).
\subsubsection{Cell-edge user}The received signal for the observed CEU-$k$ is given as follows
\begin{equation} \label{eq.rxk}
y_k = \sum_{i=1}^{B}  h_{ik}\left( \sqrt{\alpha} s_i + \sqrt{\beta} s_k \right) + n_k,
\end{equation}
where $h_{ik} \sim CN\left(0,\sigma_{ik}^2 = \acute{d}_{ik}^{-v}  \right)$ is denoted as the channel coefficient from BS-$i$ to the observed CEU-$k$, $\acute{d}_{ik}$ represents distance normalization between transmit antenna of BS-$i$ to observed CEU-$k$, and $n_k \sim (0,1)$ represents noise at CEU-$k$. By implementing the joint transmission CoMP scheme \cite{ali_compnoma}, compared to  (\ref{eq.sinrj}), the SINR value of $k$-th user is different shown below
\begin{equation} \label{eq.sinrk}
\gamma_k = \frac{\beta \sum\limits_{i=1}^{B} |h_{ik}|^2}{\alpha \sum\limits_{i=1}^{B} |h_{ik}|^2 + \frac{1}{\rho}}.
\end{equation}
\subsection{CoMP NOMA with Imperfect CSI}
In this system, the imperfect CSI is modeled by using channel estimation. Note that channel estimation error model is widely used to represent the imperfectness of CSI \cite{yang_par,wang_im,ikki_im}. Let us suppose that the estimated channel between BS-$i$ to CCU-$j$ and CEU-$k$ are represented as $\hat{h}_{ij}$ and $\hat{h}_{ik}$, respectively. The channel estimation error can be modeled as
\begin{align}
	\epsilon_{ij} = h_{ij} - \hat{h}_{ij},\label{eq.errorj}\\
	\epsilon_{ik} = h_{ik} - \hat{h}_{ik},\label{eq.errork}
\end{align}
where $\epsilon_{ij} \sim CN\left( 0,\sigma_{\epsilon_{ij}}^{2} \right)$ and $\epsilon_{ik} \sim CN\left( 0,\sigma_{\epsilon_{ik}}^{2} \right)$ are channel estimation error from BS-$i$ at CCU-$j$ and CEU-$k$, respectively. In this system, we consider $\hat{h}_{ij}$ and $\epsilon_{ij}$ as well as $\hat{h}_{ik}$ and $\epsilon_{ik}$ are independently distributed. Therefore, the distribution of $\hat{h}_{ij}$ and $\hat{h}_{ik}$ can be expressed as $CN\left( 0,\hat{\sigma}_{ij}^2 = d_{ij}^{-v} - \sigma_{\epsilon_{ij}}^{2} \right)$ and $CN\left( 0,\hat{\sigma}_{ik}^2 = \acute{d}_{ik}^{-v} - \sigma_{\epsilon_{ik}}^{2} \right)$ \cite{yang_par,wang_im,ikki_im}. 

For each user, the received signal and SINR calculation becomes different due to the channel estimation error phenomenon which can be represent as follows:
\subsubsection{Cell-center user}
The total received signals from BS-$i$ at the observed CCU-$j$ with channel estimation error is given as
\begin{equation} \label{eq.rxj_er}
\begin{aligned}
{r}_j &= \sum_{i=1}^{B}  \left(\hat{h}_{ij} + \epsilon_{ij} \right)  \left( \sqrt{\alpha} s_i + \sqrt{\beta} s_k \right) + n_j \\
&= \hat{h}_{jj}\left( \sqrt{\alpha} s_j + \sqrt{\beta} s_k \right) + \sum\limits_{\substack{i=1 \\ i \neq j}}^{B} \underbrace{\hat{h}_{ij}\left( \sqrt{\alpha} s_i + \sqrt{\beta} s_k \right)}_\textrm{interference from the other BSs}
 + \epsilon_{ij} \left( \sqrt{\alpha} s_i + \sqrt{\beta} s_k \right) + n_j.
\end{aligned}
\end{equation}
Note that even though CCU-$j$ does not perform CoMP scheme with the other BSs, CCU-$j$ still needs to estimate $B$ number of incoming signals, owing to the necessity of knowing channel information for decoding its own signal and SIC process. Similar with perfect CSI case, SIC is also applied at each CCU-$j$. Therefore, the CCU-$j$ needs to decode and remove the grouped CEU-$k$ signal before decoding its own signal with
\begin{equation} \label{eq.sinrjk_er}
\zeta_{j,k} = \frac{\beta \sum\limits_{i=1}^{B} |\hat{h}_{ij}|^2}{\sum\limits_{i=1}^{B} \left(\alpha |\hat{h}_{ij}|^2 + \sigma_{\epsilon_{ij}}^{2} \right)+\frac{1}{\rho}},
\end{equation}
where $\zeta_{j,k}$ is denoted as received SINR to decode the CEU-$k$ signal at CCU-$j$ in the presence of channel estimation error. Finally, SINR at CCU-$j$ to decode its own message with the presence of channel estimation error can be written as
\begin{equation} \label{eq.sinrj_er}
\zeta_{j} = \frac{\alpha |\hat{h}_{jj}|^2}{\alpha \sum\limits_{\substack{i=1 \\ i \neq j}}^{B}|\hat{h}_{ij}|^2 + \sum\limits_{i=1}^{B} \sigma_{\epsilon_{ij}}^{2}+\Upsilon +\frac{1}{\rho}},
\end{equation}
where $\Upsilon$ is denoted as the residual interference due to CEU-$k$ signal may not be canceled perfectly at CEU-$i$, owing to impact of estimation error in (\ref{eq.sinrjk_er}) during SIC process. The imperfect SIC may be occurred due to channel estimation error, bad pairing, and/or imperfections in the SIC process. 

\subsubsection{Cell-edge user}For the observed CEU-$k$, the total received signal in the presence of channel estimation error is given as
\begin{equation} \label{eq.rxk_er}
{r}_k = \sum_{i=1}^{B}  \left(\hat{h}_{ik} + \epsilon_{ik}\right) \left( \sqrt{\alpha} s_i + \sqrt{\beta} s_k \right) + n_k.
\end{equation}
At CEU, the multiple desired signals from all coordinated BSs are required to be detected. Consequently, the CEU also needs to estimate the incoming signal from all the coordinated BSs. Therefore, in the presence of channel estimation error, the SINR of CEU-$k$ is given as
\begin{equation} \label{eq.sinrk_er}
\zeta_{k} = \frac{\beta \sum\limits_{i=1}^{B}|\hat{h}_{ik}|^2}{\sum\limits_{i=1}^{B} \left( \alpha|\hat{h}_{ik}|^2 + \sigma_{\epsilon_{ik}}^{2} \right) +\frac{1}{\rho}}.
\end{equation}
where $\zeta_{k}$ is denoted as SINR of CEU-$k$ to decode its own signal by considering all the grouped CCUs signal as noise. 
\section{Ergodic capacity analysis} \label{sec_performance}
In this section, we derive the closed-form solutions for the exact ergodic capacity of the proposed CoMP NOMA in the presence of imperfect CSI over independent Rayleigh flat fading channel. For perfect CSI, the closed-forms can be obtained by setting up channel estimation error variance to zero. For each observed $G_k$ CoMP NOMA group, the allocated bandwidth is set with $BW$ = 1 Hz. For CCU, the closed-form is derived for two conditions, CCU with and without interference from the other BSs.
\subsection{Ergodic capacity of CCU}
Given instantaneous of received SINR in (\ref{eq.sinrj_er}), the achievable capacity of the observed CCU-$j$ is expressed as follows
\begin{equation} \label{eq.RCCU_inst}
\begin{aligned}
C_{j,\textrm{CCU}} &= \log_2(1+\zeta_{j}) \\
&= \log_2 \left( 1+ \frac{\alpha \rho |\hat{h}_{jj}|^2}{\alpha \rho \sum\limits_{\substack{i=1 \\ i \neq j}}^{B}|\hat{h}_{ij}|^2 + \rho \sum\limits_{i=1}^{B} \sigma_{\epsilon_{ij}}^{2}+\rho \Upsilon + 1 } \right) \\
&= \log_2 \left( \frac{\alpha \rho \sum\limits_{i=1}^{B}|\hat{h}_{ij}|^2 + \rho \sum\limits_{i=1}^{B} \sigma_{\epsilon_{ij}}^{2}+\rho \Upsilon +1}{\alpha \rho \sum\limits_{\substack{i=1 \\ i \neq j}}^{B}|\hat{h}_{ij}|^2 + \rho \sum\limits_{i=1}^{B} \sigma_{\epsilon_{ij}}^{2} + \rho \Upsilon + 1 } \right).
\end{aligned}
\end{equation}
Using $\log_n(x/y) = \log_n(x)-\log_n(y)$, (\ref{eq.RCCU_inst}) can be derived as
\begin{equation} \label{eq.RCCU_inst2}
\begin{aligned}
C_{j,\textrm{CCU}}
&= \log_2 \left( \alpha \rho \sum\limits_{i=1}^{B}|\hat{h}_{ij}|^2 + \underbrace{\rho \sum\limits_{i=1}^{B} \sigma_{\epsilon_{ij}}^{2} + \rho \Upsilon +1}_{a} \right) 
-
\log_2 \left(  \alpha \rho \sum\limits_{\substack{i=1 \\ i \neq j}}^{B}|\hat{h}_{ij}|^2 + \underbrace{\rho \sum\limits_{i=1}^{B} \sigma_{\epsilon_{ij}}^{2} + \rho \Upsilon + 1}_{a} \right) .
\end{aligned}
\end{equation}
Then, by using (\ref{eq.RCCU_inst2}), the ergodic capacity for CCU-$j$ is given
\begin{equation} \label{eq.Erg_CCU1}
\begin{aligned}
C_{j,\text{CCU}}^{\text{exact}} &= E\{C_{j,\textrm{CCU}}\} \\
&= \int_{0}^{\infty} \log_2 \left( x + a \right) f_{X_j}(x)dx 
- \int_{0}^{\infty} \log_2 \left( y + a \right) f_{Y_j}(y) dy,
\end{aligned}
\end{equation}
where $E$ is denoted as the expectation operator, and $a = \rho \sum\limits_{i=1}^{B} \sigma_{\epsilon_{ij}}^{2} + \rho \Upsilon +1$. Using the probability density function of $f_{X_j}(x)$ and $f_{Y_j}(y)$ which are derived in Appendix A, if more than two BSs are involved in the CoMP scheme ($B \geq 3$), (\ref{eq.Erg_CCU1}) can be written as
\begin{equation} \label{eq.Erg_CCU2}
\begin{aligned}
C_{j,\text{CCU}}^{\text{exact}}
&= \int_{0}^{\infty} \log_2 \left( x + a \right) \left( \sum\limits_{i=1}^{B} f_{X_{ij}}(x) \prod\limits_{\substack{h=1 \\ h \neq i}}^{B} \frac{k_{hj}}{k_{hj} - k_{ij}} \right) dx 
  - \int_{0}^{\infty} \log_2 \left( y + a \right) \left( \sum\limits_{\substack{i=1 \\ i \neq j}}^{B} f_{Y_{ij}}(y) \prod\limits_{\substack{h=1 \\ h \neq i \\ h \neq j}}^{B} \frac{k_{hj}}{k_{hj} - k_{ij}} \right) dy. \\
\end{aligned}
\end{equation}
Substituting (\ref{eq.pdf_xi}) and (\ref{eq.pdf_yi}) to $f_{X_{ij}}(x)$ and $f_{Y_{ij}}(y)$ in (\ref{eq.Erg_CCU2}), the $C_{j,\text{CCU}}^{\text{exact}}$ can be written as
\begin{equation} \label{eq.Erg_CCU3}
\begin{aligned}
C_{j,\text{CCU}}^{\text{exact}}
&= \int_{0}^{\infty} \log_2 \left( x + a \right)  
 \times \left( \sum\limits_{i=1}^{B} k_{ij} \exp(-k_{ij} x) \prod\limits_{\substack{h=1 \\ h \neq i}}^{B} \frac{k_{hj}}{k_{hj} - k_{ij}} \right) dx \\
& \  \ - \int_{0}^{\infty} \log_2 \left( y + a \right) 
 \times \left( \sum\limits_{\substack{i=1 \\ i \neq j}}^{B} k_{ij} \exp(-k_{ij} y) \prod\limits_{\substack{h=1 \\ h \neq i \\ h \neq j}}^{B} \frac{k_{hj}}{k_{hj} - k_{ij}} \right) dy .
\end{aligned}
\end{equation}
Then, by using  equation $\int_{0}^{\infty} \exp(- \mu x) \ln{(\beta+x)}dx = \frac{1}{\mu} \left[ \ln(\beta)-\exp(\beta \mu) \text{Ei}{(-\beta \mu)} \right]$ \cite[eq.(4.337.1)]{integral} and $\log_2 (x)=\frac{\ln(x)}{\ln(2)}$, the final expression of $C_{j,\text{CCU}}^{\text{exact}}$ with $B \geq 3$ can be written
\begin{equation} \label{eq.Erg_CCU4}
\begin{aligned}
C_{j,\text{CCU}}^{\text{exact}}
&= \frac{1}{\ln(2)} \sum\limits _{i=1}^{B} \left( \ln(a) - \exp(ak_{ij}) \textrm{Ei}(-ak_{ij}) \right) 
 \times \prod\limits_{\substack{h=1 \\ h \neq i}}^{B} \frac{k_{hj}}{k_{hj} - k_{ij}} \\
& \ \ -\frac{1}{\ln(2)} \sum\limits _{\substack {i=1 \\ i \neq j}}^{B} \left( \ln(a) - \exp(ak_{ij}) \textrm{Ei}(-ak_{ij}) \right) 
 \times \prod\limits_{\substack{h=1 \\ h \neq i \\ h \neq j}}^{B} \frac{k_{hj}}{k_{hj} - k_{ij}},  & \mbox{$B \geq 3$}.
\end{aligned}
\end{equation}
%
%
%
In case only two BSs are involved in the CoMP scheme, the $f_{Y_j}(y)$ follows (\ref{eq.sumpdfy_j2}) in Appendix A. Therefore, $(\ref{eq.Erg_CCU1})$ is derived as
\begin{equation} \label{eq.Erg_CCU6}
\begin{aligned}
C_{j,\text{CCU}}^{\text{exact}}
&= \int_{0}^{\infty} \log_2 \left( x + a \right) \left( \sum\limits_{i=1}^{2} f_{X_{ij}}(x) \prod\limits_{\substack{h=1 \\ h \neq i}}^{2} \frac{k_{hj}}{k_{hj} - k_{ij}} \right) dx \\
& \  \ - \int_{0}^{\infty} \log_2 \left( y + a \right) k_{ij} \exp(-k_{ij} y) dy, \ \ \ \ \ \ \mbox{$B=2$}.
\end{aligned}
\end{equation}
%
%
%
Let $W_{j} \triangleq \frac{\rho \alpha |\hat{h}_{jj}|^2}{1+\rho \sum\limits_{i=1}^{B} \sigma_{\epsilon_{ij}}^{2} + \rho \Upsilon}$ , if the CEU is paired with CCU without interference from the other BSs, $\sum\limits_{i=1, i \neq j}^{B}|\hat{h}_{ij}|^2 = 0$, the capacity of CCU can be written 
\begin{equation} \label{eq.Erg_CCU_noint1}
\begin{aligned}
C_{j,\text{CCU}}^{\text{exact}} &= \int_{0}^{\infty} \log_2 \left( w + 1 \right) f_{W_{jj}}(w)dw \\
&= \int_{0}^{\infty} \log_2 \left( w + 1 \right) n_{jj} \exp(-n_{jj}w)dw , \\
\end{aligned} 
\end{equation}
where $n_{jj}=\frac{1+\rho \sum\limits_{i=1}^{B} \sigma_{\epsilon_{ij}}^{2} + \rho \Upsilon}{\rho \alpha \hat{\lambda}_{jj}}$, and $\hat{\lambda}_{jj}$ represents the mean of the $j$-th exponential random variable (RV) of $W_j$. Then, the exact ergodic capacity of CCU if the interference from the other BSs is not available, can be written as
\begin{equation} \label{eq.Erg_CCU_noint2}
\begin{aligned}
C_{j,\text{CCU}}^{\text{exact}}&=-\frac{\textrm{Ei}(-n_{jj})\exp(n_{jj})}{\textrm{ln}(2)} & \mbox{$B \geq 2$,}
\end{aligned} 
\end{equation}
where $\textrm{Ei}(.)$ represents the exponential integral function.
\subsection{Ergodic capacity of CEU}
The CEU applies the CoMP scheme which involve $B$ number of BSs to improve the spectral efficiency. To accomplish, CEU also needs to estimate $B$ number of the incoming signals, which be able to increase the error estimation possibility. Given the instantaneous received SNR in (\ref{eq.sinrk_er}), the instantaneous capacity of CEU-$k$ can be written
\begin{equation} \label{eq.RCEU_inst}
\begin{aligned}
C_{j,\text{CEU}} &= \log_2(1+\zeta_{k}) \\
&= \log_2 \left( 1+\frac{\beta \sum\limits_{i=1}^{B}|\hat{h}_{ik}|^2}{\sum\limits_{i=1}^{B} \left( \alpha|\hat{h}_{ik}|^2 + \sigma_{\epsilon_{ik}}^{2} \right) +\frac{1}{\rho}} \right) \\ 
&= \log_2 \left( \frac{\rho \sum\limits_{i=1}^{B}|\hat{h}_{ik}|^2 + \rho \sum\limits_{i=1}^{B} \sigma_{\epsilon_{ik}}^{2} + 1}{\alpha \rho \sum\limits_{i=1}^{B} |\hat{h}_{ik}|^2 + \rho \sum\limits_{i=1}^{B} \sigma_{\epsilon_{ik}}^{2} +1} \right).
\end{aligned}
\end{equation}
Then, by using $\log_n(\frac{x}{y}) =  \log_n(x) - \log_n(y)$, (\ref{eq.RCEU_inst}) can be expressed as
\begin{equation} \label{eq.RCEU_inst2}
\begin{aligned}
C_{j,\text{CEU}}
&= \log_2 \left( \rho \sum\limits_{i=1}^{B}|\hat{h}_{ik}|^2 + \underbrace{\rho \sum\limits_{i=1}^{B} \sigma_{\epsilon_{ik}}^{2} + 1}_{b} \right) 
 -\log_2\left(\alpha \rho \sum\limits_{i=1}^{B} |\hat{h}_{ik}|^2 + \underbrace{\rho \sum\limits_{i=1}^{B} \sigma_{\epsilon_{ik}}^{2} +1}_{b} \right).
\end{aligned}
\end{equation}
Using (\ref{eq.RCEU_inst2}), the ergodic capacity of observed CEU-$k$ can be obtained as
\begin{equation} \label{eq.Erg_CEU1}
\begin{aligned}
C_{k,\text{CEU}}^{\text{exact}} &= E\{C_k\} \\
&= \int_{0}^{\infty} \log_2 \left( x + b \right) f_{X_k}(x)dx 
 -\int_{0}^{\infty} \log_2 \left( y + b \right) f_{Y_k}(y) dy,
\end{aligned}
\end{equation}
where $b=\rho \sum_{i=1}^{B} \sigma_{\epsilon_{ik}}^{2} +1$. By substituting $f_{X_k}(x)$ and $f_{Y_k}(y)$ with (\ref{eq.sumpdfx_k}) and (\ref{eq.sumpdfy_k}) in Appendix A, (\ref{eq.Erg_CEU1}) is given as 
\begin{equation} \label{eq.Erg_CEU2}
\begin{aligned}
C_{k,\text{CEU}}^{\text{exact}}
&= \int_{0}^{\infty} \log_2 \left( x + b \right) \sum\limits_{i=1}^{B} f_{X_{ik}}(x) \prod\limits_{\substack{h=1 \\ h \neq i}}^{B} \frac{l_{hk}}{l_{hk} - l_{ik}} dx \\
& \ -\int_{0}^{\infty} \log_2 \left( y + b \right) \sum\limits_{i=1}^{B} f_{Y_{ik}}(y) \prod\limits_{\substack{h=1 \\ h \neq i}}^{B} \frac{m_{hk}}{m_{hk} - m_{ik}}  dy.
\end{aligned}
\end{equation}
Note that $f_{X_{ik}}(x)$ and $f_{Y_{ik}}(y)$ can be obtained from (\ref{eq.pdf_xk}) and (\ref{eq.pdf_yk}). Therefore, (\ref{eq.Erg_CEU2}) is written as
\begin{equation} \label{eq.Erg_CEU2_v2}
\begin{aligned}
C_{k,\text{CEU}}^{\text{exact}}
&= \int_{0}^{\infty} \log_2 \left( x + b \right)
 \times \sum\limits_{i=1}^{B} l_{ik} \exp(-l_{ik} x) \prod\limits_{\substack{h=1 \\ h \neq i}}^{B} \frac{l_{hk}}{l_{hk} - l_{ik}} dx \\
& \ -\int_{0}^{\infty} \log_2 \left( y + b \right)
 \times \sum\limits_{i=1}^{B} m_{ik} \exp(-m_{ik} y) \prod\limits_{\substack{h=1 \\ h \neq i}}^{B} \frac{m_{hk}}{m_{hk} - m_{ik}} dy,
\end{aligned}
\end{equation}
Finally, by using similar approach in (\ref{eq.Erg_CCU2})-(\ref{eq.Erg_CCU4}), the closed-form for exact ergodic capacity of CEU-$k$ is given as
\begin{equation} \label{eq.Erg_CEU3}
\begin{aligned}
C_{k,\text{CEU}}^{\text{exact}}
&= \frac{1}{\ln(2)} \sum\limits _{i=1}^{B} \left( \ln(b) - \exp(bl_{ik}) \textrm{Ei}(-bl_{ik}) \right) 
 \times \prod\limits_{\substack{h=1 \\ h \neq i}}^{B} \frac{l_{hk}}{l_{hk} - l_{ik}} \\
& \ -\frac{1}{\ln(2)} \sum\limits _{i=1}^{B} \left( \ln(b - \exp(bm_{ik}) \textrm{Ei}(-bm_{ik}) \right)  
 \times \prod\limits_{\substack{h=1 \\ h \neq i}}^{B} \frac{m_{hk}}{m_{hk} - m_{ik}}, & \mbox{$B \geq 2$.} \\
\end{aligned}
\end{equation}
%
\subsection{Sum capacity} 
The $k$-th group of CoMP NOMA, $G_k$, has total $B+1$ users consisting of CEU-$k$ and CCU-$j$ users. Therefore, the total capacity for each observed CoMP NOMA group, $G_k$, can be expressed by:
\begin{equation} \label{eq.Erg_sum}
\begin{aligned}
C_{G_k,\text{sum}}^{\text{exact}} = C_{k,\text{CEU}}^{\text{exact}} + \sum\limits_{j=1}^{B} C_{j,\text{CCU}}^{\text{exact}}.
\end{aligned}
\end{equation}
The conventional CoMP OMA scheme is considered for comparison. However, in the OMA scheme, the available resource is divided exclusively and equally to the involved user. Therefore, the interference from the other BSs is not considered in each user for this scheme.
\begin{figure}[!t]
	\centering
	\includegraphics [width=0.4\textwidth]{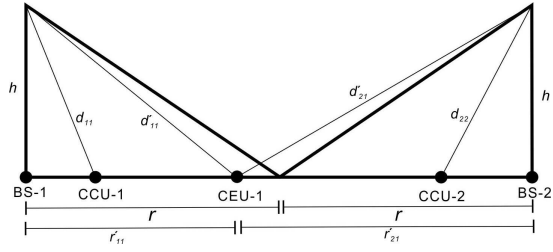} \\
	\caption {Distance representation of CoMP NOMA with study case $B=2$. $r_{11}=0.45$, $r_{22}=0.5$, and $\acute{r}_{11}=0.9$.}
	\label{fig.bs2model}
\end{figure}
\begin{figure}[!t]
	\centering
	\includegraphics [width=0.4\textwidth]{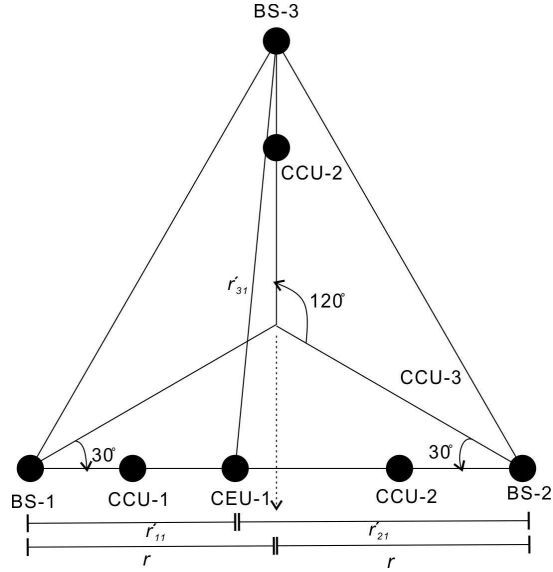} \\
	\caption {2D distance representation of CoMP NOMA with study case $B=3$ within a group G-$1$. $r_{11}=0.45$, $r_{22}=0.5$, $r_{33}=0.55$, $\acute{r}_{11}=0.9$, $\acute{r}_{21}=2r-\acute{r}_{11}$, and $\acute{r}_{31}=\sqrt{(2r)^{2}+\acute{r}^2_{11}-2(2r)\acute{r}_{11}\cos(\frac{\pi}{3})}$.}
	\label{fig.bs3model}
\end{figure}

\section{Results and discussion} \label{sec_result}
\begin{figure}[!t]
	\centering
	\includegraphics [width=0.5\textwidth]{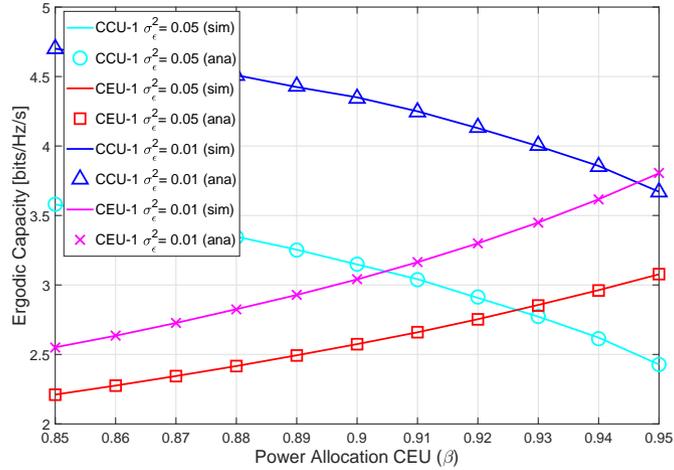} \\
	\caption {Impact of power allocation factor to CCU-1 and CEU-1 CoMP NOMA with imperfect CSI (Case I). $\Upsilon=-25$ dB and $\rho = 20$ dB.}
	\label{fig.pa}
\end{figure}

In this section, the detailed results of CoMP NOMA, with perfect and imperfect CSI, is presented and discussed. Both numerical and analytical results are provided to evaluate the performance of the proposed model. The same results between the simulation and analytical results prove the correctness of the performance analysis. The pairing scheme is conducted following two cases. Firstly, Case I, CEU-$k$ is paired among CCU-$j$, which also receives the signal from the other BSs. In this case, CCU-$j$ will consider incoming signal aside from its serving BS as interference. Otherwise, for Case II, CEU has a good pairing, so that only CEU receives the signal from the other BSs. For general parameters, we set the maximum radius model cell with distance normalization as $r=1$, the imperfectness of SIC as $\Upsilon=-25$ dB, and the height of BS as $h=0.05$. In the study case with $B=2$, we assume that BSs, CCU, and CEU in each cell are located in a parallel line as modeled in Figure \ref{fig.bs2model}. In case $B=3$, the distance model is represented in Figure \ref{fig.bs3model}, where all BSs is connected and create equilateral triangle line between them.  The distance between BS-$i$ to its served CCU-$j$ location is represented as $r_{jj}$ and to CEU-$k$ is represented as $\acute{r}_{ik}$, respectively. Then, the distance normalization between transmit antenna of BS and each user can be calculated by using the trigonometry and cosines laws, following Figs \ref{fig.bs2model} and \ref{fig.bs3model}. Note that the number of BSs can be further extended ($B \geq 2$), as long as the distance normalization between transmit antenna of BS and each user is obtained. Then, the value of path loss exponent $v$ is set as 4. We also consider the power allocation factor for CCU and CEU are $\alpha=0.05$ and $\beta=0.95$, respectively, excluding for Figure \ref{fig.pa}. For all the variance of the error estimation parameters $\sigma_{\epsilon_{ij}}^2$ and  $\sigma_{\epsilon_{ik}}^2$, we set with same value as $\sigma_{\epsilon}^2$. Moreover, the single CoMP NOMA group is assumed for sum ergodic capacity analysis. For multi-grouping scheme, it can be extended based on user pairing or grouping scheme in \cite{basit1,murti_cqi,basit_vp}.
\begin{figure}[!t]
	\centering
	\includegraphics [width=0.5\textwidth]{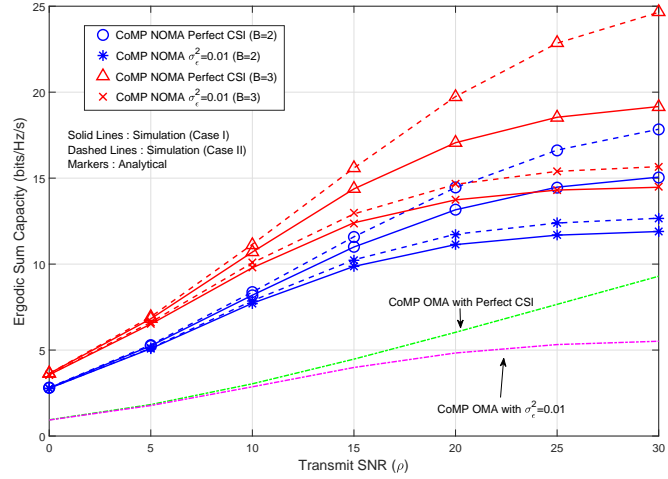} \\
	\caption {Ergodic sum capacity comparison between CoMP OMA and CoMP NOMA (Case I and Case II) with perfect and imperfect CSI. $\beta=0.95$ and $\Upsilon=-25$ dB.}
	\label{fig.sum_comparison}
\end{figure}
\begin{figure}[!t]
	\centering
	\includegraphics [width=0.5\textwidth]{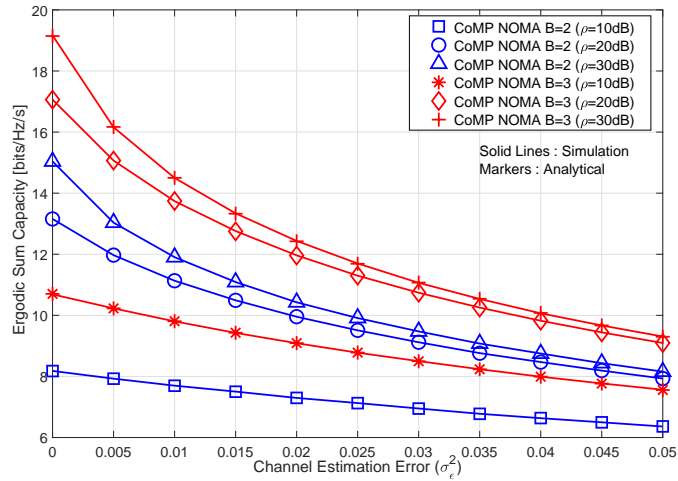} \\
	\caption {Ergodic sum capacity comparison between CoMP NOMA in the study case of $B = 2$ and $B=3$ (Case I), with imperfect CSI. $\beta=0.95$ and $\Upsilon=-25$ dB.}
	\label{fig.impact_sum}
\end{figure}

In Figure \ref{fig.sum_comparison}, we present the ergodic sum capacity comparison between CoMP OMA and CoMP NOMA with both perfect and imperfect CSI with channel estimation error. We also evaluate ergodic sum capacity in case the number of coordinated BSs increase for CoMP NOMA. In CoMP NOMA, the allocated bandwidth is shared to all the involved users within the CoMP NOMA group while the allocated bandwidth in CoMP OMA is divided equally to each user for maintaining the orthogonality. Figure \ref{fig.sum_comparison} shows that CoMP NOMA can improve more than CoMP OMA for both perfect and imperfect CSI conditions, even though CoMP NOMA also suffers from other BSs interference. The interference can be occurred at CCU in case the other BSs signals are received by CCU, owing to the NOMA scheme exploitation in the CoMP system, which allows multiple users to utilize the same frequency resource. Furthermore, if interference from the other BSs to all grouped CCUs is neglected or not occurred, further considerable capacity gain can be obtained. This condition can be achieved by employing a suitable user grouping scheme between CCU and CEU, thus, the interference can be avoided. 

Even though CoMP NOMA with two BSs can improve capacity, further capacity improvement can be obtained by increasing the number of coordinated BSs. As shown in Figure \ref{fig.sum_comparison}, by adding the number of coordinated BSs from 2 to 3, the ergodic sum capacity improves considerably for both perfect CSI and channel estimation error cases. In CoMP NOMA with all the grouped CCUs with interference from the other BSs, it is proven that CoMP NOMA ($B = 3$) still outperforms CoMP NOMA ($B = 2$) around 4 bits/Hz/s and 2.6 bits/Hz/s for perfect CSI and channel estimation error ($\sigma_{\epsilon}^2=0.001$) at $\rho= 25$ dB. A similar trend also occurs for all CCUs without interference from the other BSs. It means that the sum capacity still can be improved by increasing the number of coordinated BSs in CoMP NOMA, even though CCU and CEU also experience $B$ numbers of channel estimation error.
\begin{figure}[!t]
	\centering
	\includegraphics [width=0.5\textwidth]{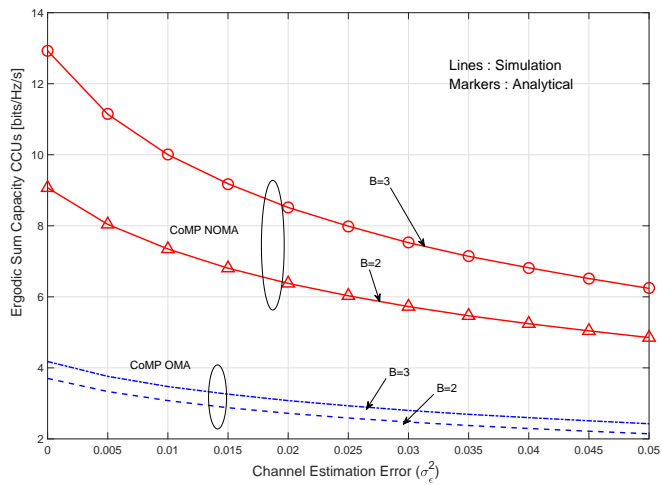} \\
	\caption {Ergodic sum capacity of CCUs comparison between CoMP OMA and CoMP NOMA (Case I) with imperfect CSI. $\beta=0.95$, $\Upsilon=-25$ dB, and $\rho = 20$ dB.}
	\label{fig.sum_ccu}
\end{figure}
\begin{figure}[!t]
	\centering
	\includegraphics [width=0.5\textwidth]{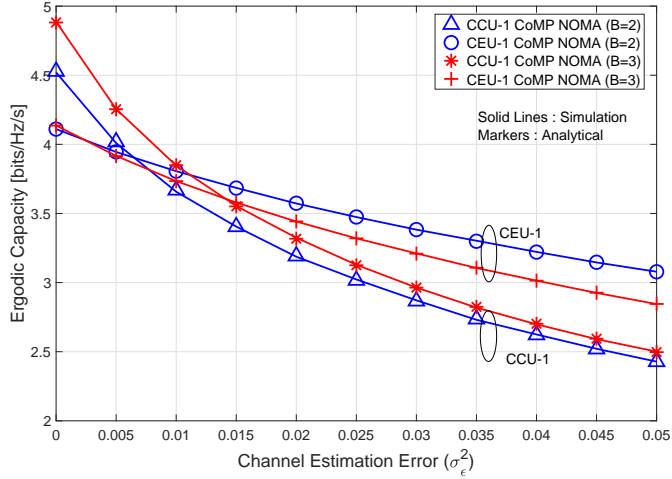} \\
	\caption {Ergodic capacity comparison CCU-1 and CEU-1 CoMP NOMA ($B=2$ and $B=3$) with imperfect CSI (Case I). $\beta=0.95$, $\Upsilon=-25$ dB, and $\rho = 20$ dB.}
	\label{fig.impact_user_bs}
\end{figure}

In Figure \ref{fig.impact_sum}, the ergodic sum capacity comparison for CoMP NOMA with $B=2$ and $B=3$ is analyzed including impact of the presence of channel estimation error over various transmit SNR values. Overall, CoMP NOMA with $B= 3$ outperforms CoMP NOMA with $B = 2$ over all channel estimation error and transmit SNR condition even though increasing more coordinated BS can lead to additional interference and channel estimation error. As discussed in (\ref{eq.RCCU_inst}) and (\ref{eq.RCEU_inst}), by adding the number of coordinated BS, the additional interference also affects CCU and CEU. Moreover, CCU and CEU must estimate the additional incoming signal that can increase error estimation effect. Figure \ref{fig.impact_sum} shows that channel estimation error provides considerable capacity degradation at high SNR compared to low SNR. For $B=2$, the ergodic sum capacity is degraded to around 3.4 bits/Hz/s at $\rho=30$ and only 1.33 bits/Hz/s at $\rho=10$ from $\sigma_{\epsilon}^2=0.01$ to $\sigma_{\epsilon}^2=0.05$. However, by exploiting the non-orthogonal channel of NOMA scheme, further sum capacity improvement still can be achieved with an increase of coordinated BSs throughout all channel estimation error parameters and $\rho$. 

%
%
%
In Figure \ref{fig.sum_ccu}, the ergodic sum capacity of CCUs is evaluated in the presence of channel estimation error. In CoMP OMA, CEU is exclusively allocated frequency resource that can not be used by the other users. Otherwise, CoMP NOMA allows CEU to share its resource with CCU.  Therefore, as shown in Figure \ref{fig.sum_ccu}, CCUs CoMP NOMA has higher sum capacity than CCUs CoMP OMA. Moreover, by adding coordinated BS to the system, the ergodic sum capacity of CCUs also increase significantly. At $\sigma_{\epsilon}^2=0.04$, CoMP NOMA with $B=3$ obtain the capacity gain around 1.5 bits/Hz/s, and 4.2 bits/Hz/s from CoMP NOMA with $B=2$ and CoMP OMA with $B=3$.

In Figure \ref{fig.impact_user_bs}, the ergodic capacity of a single CCU and CEU are compared and analyzed along with the impact of channel estimation error and increasing the number of coordinated BSs. In this result, we only consider CCU-1 and CEU-1 as analysis representation of CCU and CEU, which means sum ergodic capacity of CCUs or CEUs is not considered. This representation is also used for Figure \ref{fig.pa}. As shown in Figure \ref{fig.impact_user_bs}, the CCU-1 CoMP NOMA with $B=3$ outperforms CCU-1 CoMP NOMA with $B=2$ at whole channel estimation error condition, even though CCU-1 with $B=3$ requires estimating more incoming signals. As discussed in (\ref{eq.sinrk_er}), by increasing the number of coordinated BS, CEU needs to share the available resource with all paired CCUs, causing the increase of intra-pair interference which lead to capacity degradation. Conversely, CEU-1 with $B=2$ has higher capacity than $B=3$ because CEU with $B=2$ receives less intra-pair interference than CEU with $B=3$. Figure \ref{fig.impact_user_bs} also shows that CCU-1 experiences higher capacity degradation ($\sigma^{2}_{\epsilon} = 0$ to $\sigma^{2}_{\epsilon} = 0.05$) due to channel estimation error compared to CEU-1 for both CoMP NOMA with $B=2$ and $B=3$. The main reason of this phenomenon is that CCU acts as non-CoMP users, while CEU acts as CoMP user. Therefore, CEU can obtain benefit from all incoming signal, whereas CCU suffers from interference from all incoming signals from the other BSs.


Finally, the impact of power allocation factor toward CCU and CEU CoMP NOMA with the presence of channel estimation error, is evaluated in Figure \ref{fig.pa}. For $\sigma_{\epsilon}^2=0.05$, CCU-1 has higher capacity than CEU-1, if the CEU-1 power allocation $\beta$ is lower than 0.93. Otherwise, the ergodic capacity overwhelms CCU-1. It is clearly shown that CEU needs to be allocated with much higher power than CCU ($\beta >> \alpha$) in to maintain their capacity simultaneously. 
\section{Conclusion} \label{sec_conclusion}
In this paper, we have evaluated the proposed CoMP NOMA with perfect and imperfect CSI. Further, CCU and CEU have been analyzed in two types of scenarios, respect to consideration of the interference from the other BSs. The closed-form solutions for the exact ergodic capacity of CCU, CEU, and their sum have been derived as well. The result shows that CoMP NOMA outperforms the CoMP OMA especially if CEU are grouped among CCUs without any interference from the other BSs. It is also shown that the error estimation can lead to capacity degradation for both CCU and CEU of CoMP NOMA and CoMP OMA. In CoMP NOMA, the impact of channel estimation error is less significant at CEU compared to CCU due to utilization of incoming signals from all coordinated BSs. In addition, the capacity of CoMP NOMA can be further enhanced by increasing the number of coordinated BSs for both perfect and imperfect CSI conditions. It is confirmed that CoMP NOMA with $B=3$ considerably improves capacity compared with CoMP NOMA with $B=2$. Power allocation factor also plays an important role to CoMP NOMA. The interference analysis of proposed system shows that CEU requires higher allocated power than CCU to maintain the capacity performance. 

In future, developing an optimization framework that jointly optimize power allocation and user grouping scheme of CoMP NOMA with imperfect CSI will also be essential issues to further maximize the capacity.
\section*{Acknowledgment}
This work was supported by the National Research Foundation of Korea (NRF) grant funded by the Korea government (MSIP; Ministry of Science, ICT \& Future Planning) (2015R1D1A1A01061075).

\section*{Appendix A}
\subsection*{Derivation of $f_{X_j}(x)$ and $f_{Y_j}(y)$ for ergodic capacity of CCU}
Let us suppose $X_j \triangleq \alpha \rho \sum_{i=1}^{B}|\hat{h}_{ij}|^2 $ and $Y_j \triangleq  \alpha \rho \sum_{i=1,i \neq j}^{B}|\hat{h}_{ij}|^2$ . Then, the PDF of $X_j$ can be calculated by using sum of total $B$ i.i.d exponential RVs with different parameters and $Y_j$ can be also calculated by using sum of total $B-1$ i.i.d exponential RVs with different parameters. The parameters of exponential RVs are assumed to be different due to the distance for each BS-$i$ with CCU-$j$ is different, which cause the channel gain variance is also different for each $\hat{h}_{ij}$. For each exponential RV, the PDF can be expressed as
\begin{equation} \label{eq.pdf_xi}
\begin{aligned}
f_{X_{ij}}(x) &= \frac{d(F_{X_{ij}}(x))}{dx} = \frac{d(1- \exp(-k_{ij} x))}{dx} \\
&= k_{ij} \exp(-k_{ij} x),
\end{aligned}
\end{equation}
\begin{equation} \label{eq.pdf_yi}
\begin{aligned}
f_{Y_{ij}}(y) &= \frac{d(F_{Y_{ij}}(y))}{dy} = \frac{d(1- \exp(-k_{ij} y))}{dy} \\
&= k_{ij} \exp(-k_{ij} y),
\end{aligned}
\end{equation}
where $k_{ij}$ represents each parameter of exponential RVs, which is written as
\begin{equation} \label{eq.ki}
k_{ij} = \frac{1}{\alpha \rho \lambda_{ij}},
\end{equation}
and $\lambda_{ij}$ represents the mean of the $i$-th exponential RVs of $X_j$ and $Y_j$. Then, by using sum exponential random variable with different variable concept in \cite[eq. 7]{bibi_exp}, the PDF of $X_j$ and $Y_j$ can be derived as
\begin{equation} \label{eq.sumpdfx_j}
\begin{aligned}
f_{X_{j}}(x) &= f_{X_{ij}+X_{(i+1)j}+...+X_{Bj}}(x)  \\
&=  \sum\limits_{i=1}^{B} f_{X_{ij}}(x) \prod\limits_{\substack{h=1 \\ h \neq i}}^{B} \frac{k_{hj}}{k_{hj} - k_{ij}} , & \mbox{$B \geq 2$}.
\end{aligned}
\end{equation}
\begin{equation} \label{eq.sumpdfy_j}
\begin{aligned}
f_{Y_{j}}(y) &= f_{Y_{ij}+Y_{(i+1)j}+...+Y_{Bj}}(y), & \mbox{$i \neq j$}  \\
&=  \sum\limits_{\substack{i=1 \\ i \neq j}}^{B}  f_{Y_{ij}}(y) \prod\limits_{\substack{h=1 \\ h \neq i \\ h \neq j}}^{B} \frac{k_{hj}}{k_{hj} - k_{ij}}, & \mbox{$B \geq 3$}.
\end{aligned}
\end{equation}
However, if $B=2$, the PDF of $f_{Y_j}$ only consists with a single exponential RV. Therefore, the PDF is given as
\begin{equation} \label{eq.sumpdfy_j2}
\begin{aligned}
f_{Y_{j}}(y)
&=  k_{ij} \exp{(-k_{ij}y)} , & \mbox{$B=2$ and $i \neq j$.}
\end{aligned}
\end{equation}
\subsection*{Derivation of $f_{X_k}(x)$ and $f_{Y_k}(y)$ for ergodic capacity of CEU}
Let us suppose $X_k \triangleq \rho \sum_{i=1}^{B}|\hat{h}_{ik}|^2$ and $Y_k \triangleq \alpha \rho \sum_{i=1}^{B}|\hat{h}_{ik}|^2$. $X_k$ and $Y_k$ are sum of total $B$ i.i.d exponential RVs with different mean parameter. Using similar approach in (\ref{eq.pdf_xi}) and (\ref{eq.pdf_yi}), the PDF of each element $X_k$ and $Y_k$ is given
\begin{equation} \label{eq.pdf_xk}
\begin{aligned}
f_{X_{ik}}(x) &= \frac{d(F_{X_{ik}}(x))}{dx} = \frac{d(1- \exp(-l_{ik} x))}{dx} \\
&= l_{ik} \exp(-l_{ik} x),
\end{aligned}
\end{equation}
\begin{equation} \label{eq.pdf_yk}
\begin{aligned}
f_{Y_{ik}}(y) &= \frac{d(F_{Y_{ij}}(y))}{dy} = \frac{d(1- \exp(-m_{ik} y))}{dy} \\
&= m_{ik} \exp(-m_{ik} y),
\end{aligned}
\end{equation}
where $l_{ik}$ and $m_{ik}$ represent each parameter of element exponential RVs of $X_k$ and $Y_k$, which are given by
\begin{equation} \label{eq.lk}
l_{ik} = \frac{1}{\rho \lambda_{ik}},
\end{equation}
\begin{equation} \label{eq.mk}
m_{ik} = \frac{1}{\alpha \rho \lambda_{ik}},
\end{equation}
where $\lambda_{ik}$ represents the mean of the $i$-th exponential RVs of $X_k$ and $Y_k$. Then, following equation in \cite[eq. 7]{bibi_exp}, the PDF of $X_k$ and $Y_k$ can be written by
\begin{equation} \label{eq.sumpdfx_k}
\begin{aligned}
f_{X_{k}}(x) &= f_{X_{ik}+X_{(i+1)k}+...+X_{Bk}}(x)  \\
&=  \sum\limits_{i=1}^{B} f_{X_{ik}}(x) \prod\limits_{\substack{h=1 \\ h \neq i}}^{B} \frac{l_{hk}}{l_{hk} - l_{ik}}, & \mbox{$B \geq 2$}	
\end{aligned}
\end{equation}
\begin{equation} \label{eq.sumpdfy_k}
\begin{aligned}
f_{Y_{k}}(y) &= f_{Y_{ik}+Y_{(i+1)k}+...+Y_{Bk}}(y)  \\
&=  \sum\limits_{i=1}^{B} f_{Y_{ik}}(y) \prod\limits_{\substack{h=1 \\ h \neq i}}^{B} \frac{m_{hk}}{m_{hk} - m_{ik}}, & \mbox{$B \geq 2$}.
\end{aligned}
\end{equation}

\bibliography{ref_ce}

\section*{Author Biography}
\begin{biography}{}{\textbf{Fahri Wisnu Murti} received his B.S. from Telkom University, Bandung, Indonesia (2010-2014) and completed his M.S. at WENS Lab, Department of IT Convergence Eng., Kumoh National Institute of Technology,  South Korea (2016-2018). He was a student internship at Huawei Technologies, Shenzhen \& Beijing, PR China (2013); and network engineer at Nokia Networks, Jakarta, Indonesia (2014-2016). He was a research assistant in the School of Computer Science and Statistics, Trinity College Dublin, Ireland (2019). Currently, He is pursuing a Ph.D. degree in the Faculty of Information Technology and Electrical Engineering (ITEE) in the University of Oulu, Finland. His major research interests include optimization and learning techniques for intelligent wireless networks.}
\end{biography}
\begin{biography}{}{\textbf{Rahmat Faddli Siregar} received his B.S. degree in telecommunication engineering from Telkom University, Bandung, Indonesia, in 2016. He further accomplished his M.S degree in Kumoh National Institute of Technology (KIT), Gumi, South Korea, in 2018. During his M.S study, he also joined and involved several projects in Wireless and Emerging Network System (WENS) laboratory. Currently, he is pursuing Ph.D degree in Faculty of Information Technology and Electrical Engineering (ITEE) in University of Oulu, Finland and involves in 6G flagship project under Centre for Wireless Communication. His major research interests include information theory, wireless communication optimization, multiple input multiple output (MIMO), millimeter wave, non-orthogonal multiple access (NOMA), index modulation, etc.}
\end{biography}
\begin{biography}{}{\textbf{Muhammad Royyan} received his M.E. from Kumoh National Institute of Technology, South Korea, in 2018 and received his B.E from Telkom University, Indonesia, in 2015. He worked as Junior System Engineer from 2014 until 2015 at Indonesian Aerospace.  He is currently a Doctoral Candidate in the Department of Signal Processing and Acoustics at Aalto University, Finland. His research interests are in networked control systems, distributed systems, and large-scale data analysis.}
\end{biography}
\begin{biography}{}{\textbf{Soo Young Shin}  received his Ph.D. degrees in electrical engineering and computer science from Seoul National University on 2006. He was with WiMAX Design Lab, Samsung Electronics, Suwon, South Korea from 2007 to 2010. He joined as full-time professor to School of Electronics, Kumoh National Institute of Technology, Gumi, South Korea. He is currently an Associate Professor. He was a post Doc. researcher at University of Washington, Seattle, WA, USA from 2006 to 2007. In addition, he was a visiting scholar to University of the British Columbia at 2017. His research interests include wireless communications, next generation mobile wireless broadband networks, signal processing, Internet of things, etc.}
\end{biography}
\end{document}